# A Modified Mixed Domain Method for Modeling Acoustic Wave Propagation in Strongly Heterogeneous Media

Juanjuan Gu and Yun Jing

Department of Mechanical and Aerospace Engineering, North Carolina State University,

Raleigh, NC 27695, USA

**Abstract**

In this paper, phase correction and amplitude compensation are introduced to a previously developed mixed domain method (MDM), which is only accurate for modeling wave propagation in weakly heterogeneous media. Multiple reflections are also incorporated with the one-way model to improve the accuracy. The resulting model is denoted as the modified mixed-domain method (MMDM) and is numerically evaluated for its accuracy and efficiency using three distinct cases. It is found that the MMDM is significantly more accurate than the MDM for strongly heterogeneous media, especially when the phase aberrating layer is approximately perpendicular to the acoustic beam. Additionally, a convergence study suggests that the second-order reflection is sufficient for wave modeling in lossy biological media. The method developed in this work is expected to facilitate therapeutic ultrasound for treating brain-related diseases and disorders.

## I. Introduction

Numerical modeling of acoustic wave propagation in heterogeneous media is of great importance to medical ultrasound. In therapeutic ultrasound applications, for example, numerical simulations can be used to study the phase aberration in MR-guided focused ultrasound surgery[1,2] and to improve the treatment outcome. For diagnostic ultrasound, numerical modeling has been used as an important tool for image reconstruction[3,4,5] as well as to understand the sources of image degradation in ultrasound imaging[6].

Many wave propagation algorithms that take medium heterogeneities into account have been developed. Most of these algorithms operate in the time-domain. Demi *et al.*[7] presented an iterative nonlinear contrast source method for modeling nonlinear acoustic wave propagation in media with spatially inhomogeneous attenuation. They later extended the method to more general cases[8]. Treeby *et al.*[9] developed a k-space time-domain (KSTD) method using the coupled nonlinear wave equation. Jing *et al.*[10] alternatively developed the KSTD from the Westervelt equation. Pinton *et al.*[11] studied a heterogeneous nonlinear attenuating full-wave model based on the finite-difference time-domain (FDTD) method. Frequency-domain methods have also been investigated. For example, Clement and Hynynen[12] combined the Angular Spectrum Approach (ASA) with ray theory to describe the propagation of ultrasound through randomly oriented, dissipative, layered media. Vyas and Christensen[13] modified the conventional ASA method to model linear wave propagation in inhomogeneous media. Most recently, a mixed domain method (MDM) for modeling linear/nonlinear wave propagation in dissipative, weakly heterogeneous media has been presented[14,15]. A detailed summary of modern ultrasound modeling algorithms can be found in a review paper[16].

The present paper aims to establish and validate an accuracy-efficiency balanced numerical model for simulating acoustic wave propagation in strongly heterogeneous media. Within the realm of biomedical ultrasound, this model is particularly pertinent to transcranial ultrasound, and could therefore facilitate research on high intensity focused ultrasound (HIFU) for treating brain-related diseases[17,18] as well as research on ultrasound-mediated neuromodulation[19]. This numerical model is a non-trivial extension to the previously developed MDM[14], which is a one-way model and is only accurate for weakly heterogeneous media. To extend the original MDM to modeling wave propagation in strongly heterogeneous media, phase and amplitude corrections are proposed and evaluated in this paper. The phase correction term is first theoretically derived. As the transmission coefficient due to the variation of sound speed is not considered in the original MDM, an amplitude compensation term is also proposed. Reflections are added to the one-way model to further improve the accuracy. The resulting method is denoted the modified mixed domain method (MMDM). A two-dimensional (2D) layered medium, a 2D human skull, and a three-dimensional (3D) skull-mimicking medium are studied to evaluate the accuracy of the MMDM. Results from the MATLAB toolbox k-Wave[20] are used as the benchmark for comparison and validation purposes. The reason we chose k-Wave as the benchmark is that k-Wave is a well-established toolbox for acoustic wave simulations[20] and it has been used as the benchmark in our previous study[14]. This study shows that the MMDM can markedly improve the results for strongly heterogeneous media in terms of the predicted waveform phase and amplitude, provided that the phase aberrating layer is approximately perpendicular to the ultrasound beam. While the addition of reflections can improve the accuracy of the model, it is also found that up to the second-order reflection is sufficient for obtaining converged results when sound absorption is considered, i.e., higher order reflections do not significantly improve the result. This paper is structured as follows:

Section II puts forward the phase correction, the amplitude compensation term, and the scheme for modeling the reflection. Section III systematically evaluates the MMDM by comparing its results with those of k-Wave and the MDM. Section IV discusses both the strength and weakness of the MMDM. Section V concludes the paper.

## II. Theory

### A. Governing equation

We begin with the generalized Westervelt equation[16] and it reads

$$\rho \nabla \cdot \left(\frac{1}{\rho}\nabla p\right) - \frac{1}{c^2}\frac{\partial^2 p}{\partial t^2} + \frac{\delta}{c^4}\frac{\partial^3 p}{\partial t^3} + \frac{\beta}{\rho c^4}\frac{\partial^2 p^2}{\partial t^2} = 0, \qquad (1)$$

where $p$ is the acoustic pressure, $\rho$ is the ambient density, $c$ is the speed of sound, $\delta$ is the sound diffusivity, $\delta = 2\alpha_{NP}c^3/\omega^2$ ($\alpha_{NP}$ is the attenuation coefficient in $Np/m$ and $\omega$ is the angular frequency), and $\beta$ is the nonlinearity coefficient. In the original MDM, Eq. (1) would be first transformed by applying the normalized wave field $f = p/\sqrt{\rho}$ and the equation yields[14]

$$\nabla^2 f - \frac{1}{c^2}\frac{\partial^2 f}{\partial t^2} - f\sqrt{\rho}\nabla^2\frac{1}{\sqrt{\rho}} + \frac{\delta}{c^4}\frac{\partial^3 f}{\partial t^3} + \frac{\beta}{\sqrt{\rho}c^4}\frac{\partial^2 f^2}{\partial t^2} = 0. \qquad (2)$$

The effect of density heterogeneities is taken into account by the term $\sqrt{\rho}\nabla^2\frac{1}{\sqrt{\rho}}$. In the event that the density distribution is not sufficiently smooth, which could be the case for heterogeneous media, the Laplacian term $\nabla^2\frac{1}{\sqrt{\rho}}$ will exhibit sharp discontinuities[21]. While this was not identified as a major issue for weakly heterogeneous media[14], it could render the algorithm unstable for strongly heterogeneous cases. A previous paper also discussed the adverse effect of this Laplacian term in the context of the KSTD method[22]. Consequently, the density is first assumed to be homogeneous in the governing equation. The density heterogeneity effect will be later considered via an amplitude correction term proposed in part C of this section. To reduce the spatial aliasing error[23], an absorbing boundary layer is added by introducing a frequency-independent absorption term $\gamma \frac{\partial p}{\partial t}$ to the governing equation[24], where $\gamma = \gamma_{max}/\cosh^2(\alpha n)$ ($\gamma_{max}$ is a constant, $\alpha$ is a decay factor, and $n$ denotes the distance in number of grid points from the boundary). Thus, the modified governing equation reads

$$\nabla^2 p - \frac{1}{c^2}\frac{\partial^2 p}{\partial t^2} + \frac{\delta}{c^4}\frac{\partial^3 p}{\partial t^3} + \frac{\beta}{\rho c^4}\frac{\partial^2 p^2}{\partial t^2} = \gamma \frac{\partial p}{\partial t}. \tag{3}$$

This equation is akin to the original Westervelt equation, in which the sound speed $c$ and diffusivity $\delta$ are constant with respect to the frequency. Subtracting $\frac{1}{c_0^2}\frac{\partial^2 p}{\partial t^2}$ from both sides of Eq. (3), where $c_0$ is a constant (generally taken as the minimum sound speed in the medium under study), and rearranging the resulting equation yields

$$\nabla^2 p - \frac{1}{c_0^2}\frac{\partial^2 p}{\partial t^2} = \left(\frac{1}{c^2} - \frac{1}{c_0^2}\right)\frac{\partial^2 p}{\partial t^2} - \frac{\delta}{c^4}\frac{\partial^3 p}{\partial t^3} - \frac{\beta}{\rho c^4}\frac{\partial^2 p^2}{\partial t^2} + \gamma \frac{\partial p}{\partial t}. \tag{4}$$

By performing the Fourier transform to Eq. (4) with respect to $x$, $y$ and time $t$, we arrive at an equation in the frequency domain,

$$\frac{\partial^2}{\partial z^2}\tilde{P} + K^2 \tilde{P} = F_{xy}\left\{\left[-\frac{\omega^2}{c_0^2}\left(\frac{c_0^2}{c^2} - 1\right) + \frac{i\delta\omega^3}{c^4} + i\omega\gamma\right]F_t(p)\right\} + F_{xy}\left(\frac{\beta\omega^2}{\rho c^4}F_t(p^2)\right), \tag{5}$$

where $\tilde{P}$ is the Fourier transform of $p$, $F_{xy}$ is the Fourier transform operator in $x$- and $y$-dimensions. We note that, applying the Fourier transform ($F_{xy}$) to the terms containing speed of sound, diffusivity, $\gamma$, and nonlinearity coefficient in Eq. (5) mathematically give rise to convolutions in the wave-vector domain. Our algorithm computes these convolutions by leveraging the fast Fourier transform (FFT). $F_t$ is the Fourier transform in the time domain, and $K^2 = \omega^2/c_0^2 - k_x^2 - k_y^2$ with $k_x$ and $k_y$ being the wave-numbers in $x$- and $y$- dimensions. This equation is more general than Eq. (3) in the sense that now the speed of sound and diffusivity $\delta$ can be treated as arbitrary functions of the frequency in order to take dispersion into account[25]. An implicit solution to Eq. (5) is derived from the 1-D Green's function in an integral form[25], such that

$$\tilde{P}(z) = \tilde{P}(0)e^{iKz} + \frac{e^{iKz}}{2iK}\int_0^z e^{-iKz'} M(p(z'))dz', \tag{6}$$

where

$$M(p) = F_{xy}\left\{\left[-\frac{\omega^2}{c_0^2}\left(\frac{c_0^2}{c^2}-1\right) + \frac{i\delta\omega^3}{c^4} + i\omega\gamma\right]F_t(p)\right\} + F_{xy}\left(\frac{\beta\omega^2}{\rho c^4}F_t(p^2)\right). \tag{7}$$

Equation (6) is solved by using a Simpson-like rule[26]. In this model, wave effects such as diffraction, attenuation, dispersion and nonlinearity are all considered. Additionally, density, speed of sound, attenuation coefficient, power law exponent and nonlinear coefficient can all be spatially varying functions. The Kramers-Kronig dispersion relation is applied by directly replacing the speed of sound $c$ with $c_p$ and $c_p = (1/\hat{c} + \alpha_0 \tan(\pi y/2)\omega^{y-1})^{-1}$[25], where $\hat{c}$ is the sound speed at zero frequency[27], $y$ is the power law exponent, $\alpha_0$ is the absorption in Np·MHz$^{-y}$·m$^{-1}$. The relation between $\alpha_0$ and $\alpha_{NP}$ follows a power law, i.e., $\alpha_{NP} = \alpha_0\omega^y$. This model, however, is only accurate for media with weak speed of sound contrast. As shown by our previous study[14], this model is a one-way model; it does not consider the transmission coefficient associated with the speed of sound variation. There is also an intrinsic error when computing the phase of the advancing wavefront, which grows as the speed of sound contrast increases[14]. To have a more general model that could be applied to strongly heterogeneous media, phase correction and transmission compensation will be introduced. Multiple reflections are also proposed to complement the model.

### B.    Phase correction

Considering a one-dimensional (1D) inhomogeneous medium with a speed of sound distribution that is

$$c = \begin{cases} c_1, \text{when } z \leq z_0 \\ c_2, \text{when } z > z_0 \end{cases}. \tag{8}$$

The analytical solution of the pressure at $z + \Delta z$ ($z = z_0$) without considering the transmission coefficient (only consider the phase change) is

$$(P_{z+\Delta z})_{analytical} = P_z e^{iK'\Delta z} \tag{9}$$

where $K'$ is the wave number and $K' = \omega/c_2$, $P_z$ is the wave pressure at $z$ with a frequency of $\omega$. The original MDM solution, on the other hand, is described as[14]

$$(P_{z+\Delta z})_{MDM} = P_z e^{iK\Delta z} + \frac{e^{iK\Delta z}}{2iK} \int_0^{\Delta z} e^{-iKz'}(M)P(z')\,dz', \tag{10}$$

For 1D linear lossless wave propagation, $K = \omega/c_0$ and $M = -\frac{\omega^2}{c_0^2}\left(\frac{c_0^2}{c^2} - 1\right)$. In this case, $c_0 = c_1$. It has been rigorously proven that this solution is only valid for weakly heterogeneous media[14]. To solve the integral equation in the form of $y(t_{n+1}) = y(t_n) + \int_{t_n}^{t_{n+1}} f(s, y(s))\,ds$, the Trapezoidal rule is applied and it yields

$$y_{n+1} = y_n + \frac{\Delta z}{2}[f(t_n, y_n) + f(t_{n+1}, y_{n+1})]. \tag{11}$$

Applying this to Eq. (10) leads to

$$(P_{z+\Delta z})_{MDM} = P_z e^{iK\Delta z} + \frac{e^{iK\Delta z}}{2iK}\frac{\Delta z}{2}\left[M_z P_z + M_{z+\Delta z} P_{z+\Delta z} e^{-iK\Delta z}\right]. \tag{12}$$

By rearranging Eq. (12), we have

$$(P_{z+\Delta z})_{MDM} = \frac{P_z e^{iK\Delta z} + \frac{e^{iK\Delta z}}{4iK} M_z P_z \Delta z}{1 - \frac{M_{z+\Delta z}}{4iK}\Delta z}. \tag{13}$$

To examine the exact phase error in the MDM, Eq. (13) is subtracted from Eq. (9). Rearranging the resulting equation yields

$$(P_{z+\Delta z})_{analytical} = (P_{z+\Delta z})_{MDM} + P_z e^{iK'\Delta z}\left(1 - e^{i(K-K')\Delta z}\frac{1 + \frac{M_z}{4iK}\Delta z}{1 - \frac{M_{z+\Delta z}}{4iK}\Delta z}\right). \tag{14}$$

$$P_z e^{iK'\Delta z}\left(1 - e^{i(K-K')\Delta z}\frac{1+\frac{M_z}{4iK}\Delta z}{1-\frac{M_{z+\Delta z}}{4iK}\Delta z}\right)$$ is therefore the phase correction term. Although this correction is derived based on the 1D assumption, it can be applied to more general cases with a sufficient accuracy as shown later in this paper.

### C. Amplitude compensation

The transmission coefficient due to the variation of sound speed is not considered in the original MDM, which is a significant source of error for simulations involving a large speed of sound contrast. Although the MDM could consider the transmission coefficient due to the variation of density, as stated earlier, the density heterogeneity term could introduce a singularity and render the algorithm unstable. Therefore, an amplitude compensation is introduced for addressing the density and speed of sound heterogeneities. The compensation term reads (similar to what was used in reference[13])

$$T(x,y,z) = \frac{2\rho(x,y,z+\Delta z)c(x,y,z+\Delta z)}{\rho(x,y,z)c(x,y,z)+\rho(x,y,z+\Delta z)c(x,y,z+\Delta z)}, \quad (15)$$

where $c(x,y,z)$ and $\rho(x,y,z)$ are the speed of sound and density at plane $z$, respectively; $c(x,y,z+\Delta z)$ and $\rho(x,y,z+\Delta z)$ are the speed of sound and density at plane $z+\Delta z$, respectively. Similar to Eq. (14), Eq. (15) is only exact for 1D cases[28]. To implement the phase and amplitude corrections, the second term on the right hand side of Eq. (14) is added to $\tilde{P}(z)$ in Eq. (6) during the iteration for computing the integral. After applying the inverse Fourier transform to $\tilde{P}(z)$, the amplitude is then corrected by multiplying $T(x,y,z)$.

### D. Multiple reflections

Reflections is further added in the MMDM by using the following equation[28]

$$P_{reflection} = P_{incident}(T - 1), \tag{16}$$

where $P_{incident}$ is the incident wave used for calculating the reflected wave. For example, when calculating the first-order reflection, $P_{incident}$ is the transmissive waveform at each plane, i.e., the result of the one-way MMDM. The corresponding $P_{refleciton}$ is first calculated for each layer by Eq. (16) and stored during the forward projection step. Subsequently, the entire first-order reflection field is computed by considering $P_{reflection}$ as the boundary condition and having it propagate in the backward direction. When calculating the second-order reflection, $P_{incident}$ is given by the first-order reflection wave field. By propagating the resulting $P_{reflection}$ in the forward direction, the second-order reflection field is formed. This procedure continues until the desired maximum order of reflection is reached. In general, an even-order reflection is associated with forward propagation while an odd-order reflection travels in the backward direction. The final wave field is obtained by superposing all solutions. It is noted that, again, Eq. (16) is only exact for 1D wave propagation, which is consistent with the assumption underpinning Eqs. (14) and (15). A flowchart illustrating how the corrections and reflections are implemented in the MMDM can be found in Fig. 1.

## III. Simulation Results

### A. Short tone bursts

For transient simulations, a Gaussian-modulated pulse is used and is expressed as

$$p_{source} = p_0 \exp(-t^2 f_c^2 / 2) \sin(2\pi f_c t), \tag{17}$$

where $p_0$ is the magnitude of the pulse and $f_c$ is the center frequency. Benchmark results are obtained by the MATLAB toolbox k-Wave[20]. Both spatial and temporal resolutions used in the benchmark simulations are sufficiently fine in order to obtain well-converged results (please see the end of next paragraph for the definition of "well-converged"). L2-norm errors are calculated to quantitatively analyze the accuracy of the MMDM and this error is defined as[15]

$$L2 = \frac{\|p - p_{benchmark}\|}{\|p_{benchmark}\|}, \qquad (18)$$

where $\|p\|$ is the L2-norm of the acoustic pressure.

A layered medium shown in Fig. 2(a) is first studied. In the blue region, the speed of sound is 1500 m/s and density is 1000 kg/m³; in the red region, the speed of sound is 3000 m/s and density is 2000 kg/m³, indicating a contrast of 2.0 for both acoustical properties. The transducer focal length is 68.6 mm and the transducer diameter is 34.3 mm, corresponding to an *F* number of 2.0. The transducer center frequency is 0.7 MHz and the pressure magnitude of the excitation $p_0$ is 1 Pa. The absorbing layer is enabled to minimize the spatial aliasing error. For the transient simulation, $\gamma_{max}$ is 0.6 and $\alpha$ is 0.05. Note that, a large $\gamma_{max}$ or a small $\alpha$ implies that more energy will be absorbed by the absorbing layer, thus reducing the spatial aliasing error (sometimes also referred to as the wrap-around error). On the other hand, if $\gamma_{max}$ is too large or $\alpha$ is too small, the algorithm could become unstable with a large step size or the acoustic field could become inaccurate due to excess absorption. In practice, the best $\gamma$ distribution yields near-zero values in the central area of the domain while the $\gamma$ value increases smoothly towards the edge of the domain. More details regarding setting up the absorbing layer can be found in ref. [24]. For the benchmark simulation using k-Wave, the spatial step size is 1/16 λ (λ is the wavelength in water at the center frequency). The time step size *dt* is 0.0045 μs, corresponding to a Courant-Friedrichs-Lewy (CFL) number of 0.1. The error of the k-Wave results at CFL=0.05 using the results at CFL=0.1 as the benchmark is on the order of 0.001, which is negligibly small when compared to the difference between the results of our algorithm and the benchmark results. The same criterion is also used to determine the spatial resolution for convergence.

For the MDM and MMDM simulations, the spatial step size in the *x* direction is 1/4 λ and it is 1/16 λ in the *y* direction (propagation direction). The time step size $dt$ is 0.1786 μs. Additional simulations show that smaller $dt$ does not significantly affect the result once the Nyquist sampling rate is well satisfied (i.e., $dt \leq 1/8f_c$). Waveforms recorded at the transducer focus simulated with different methods are shown and compared in Fig. 2(b). The L2-norm error is 1.2137 for the simulation with MDM and is 0.0509 for the simulation with MMDM incorporating up to the fourth-order reflection (denoted as MMDM4; MMDMn stands for the MMDM incorporating up to the nth-order reflection). When the nonlinear effect is considered, the pressure magnitude is increased from 1 Pa to 1 MPa. The nonlinearity coefficient for the whole domain is 3.6. The time-domain and frequency-domain results at the focal point are plotted in Figs. 2(c)-2(d). The L2-norm error is 1.2234 for MDM and is 0.0579 for MMDM4. In both linear and nonlinear cases, the error is reduced by a factor of over 20. In both Figs. 2(b) and 2(d), even-order reflections can be observed in the MMDM and k-Wave results, as anticipated.

A 2D human skull is then studied to further validate the MMDM and the *in silico* model is shown in Fig. 3(a). The speed of sound is between 1500 m/s and 2816.1 m/s; the density is between 1000 kg/m³ and 2588 kg/m³. The transducer focal length is 58.6 mm and its diameter is 39.3 mm, corresponding to an *F* number of 1.5. The transducer center frequency is 0.7 MHz and the pressure amplitude is 1 Pa. The spatial step size in the *x* and *y* directions are both 0.1953 mm (approximately 1/11λ) for all simulations, which is determined by the CT scan. The time step size in k-Wave is 0.0045 μs (CFL=0.065) while it is 0.1786 μs in the MDM and MMDM. The medium is considered to be lossless. The effect of acoustic absorption will be addressed in the discussion section. The absorbing layer is used to minimize the spatial aliasing error, where $\gamma_{max}$ is 0.6 and $\alpha$ is 0.05. We first compare the waveforms recorded at the focus of the transducer, which are plotted in Fig. 3(b).

The L2-norm error is 1.0716 for MDM and is 0.1120 for MMDM4. In contrast to the previous case, the even order reflections are not visible in this case as they are mixed with the primary transmissive ($0^{th}$ order) wave. The nonlinear effect is subsequently considered. The nonlinearity coefficient is 3.6 throughout the entire domain, though in principle it can be inhomogeneous in the MDM/MMDM. The pressure magnitude is increased to 1 MPa to enhance the nonlinear effect. The time-domain and frequency-domain results at the focal point of the transducer are shown in Figs. 3(c)-3(d). The L2-norm error is 1.0967 for MDM and is 0.1261 for MMDM4. In this case, the error is reduced by a factor of almost 10 for both linear and nonlinear simluations.

**B. Continuous wave beams**

The MMDM is also capable of directly modeling the acoustic field at the frequency of interest[15] since this method is intrinsically a frequency-domain method. It is shown that, compared to transient simulations, where the acoustic field of a certain frequency needs to be acquired by Fourier transform, directly operating the MDM/MMDM at the frequency of interest is orders of magnitude more computationally efficient. This, however, has only been demonstrated for linear and weakly nonlinear cases, where the couplings between the fundamental frequency and the harmonics are minimal[15]. In this study, linear wave propagation is assumed and the excitation signal in k-Wave is a continuous sinusoidal wave at 0.7 MHz with an amplitude of 1 Pa. For the same 2D skull case, the absorbing layer in MDM/MMDM is enabled where $\gamma_{max}$ is chosen as 0.5 and $\alpha$ is 0.03. Four sets of results, obtained by k-Wave, MDM, MMDM and MMDM4, are plotted in Figs. 4(a)-(d). Figures 4(e) and (f) show the pressure difference between MDM and k-Wave, and MMDM4 and k-Wave, respectively. For the region shown in Figs. 4(a)-(d) (approximately 80 mm by 80 mm), the L2-norm error is 0.3374 for MDM, 0.2839 for MMDM, and 0.1717 for MMDM4, respectively. Axial pressure distributions along $x = 0$ are also compared in Fig. 4(g) between different models.

We have also conducted 3D simulations. A 3D skull-mimicking model is generated as shown in Fig. 5, with a domain size of 42.9 mm ×42.9 mm ×42.9 mm. For the surrounding medium, the speed of sound is 1500 m/s and the density is 1000 kg/m$^3$; for the skull-like medium, the speed of sound is 2250 m/s and the density is 1500 kg/m$^3$. An unfocused circular transducer is adopted in this simulation, whose diameter is 17.14 mm. The center frequency and the pressure amplitude are 0.7 MHz and 1 Pa, respectively. The same spatial resolution is used for all methods and it is 1/10 λ. For the benchmark simulation using k-Wave, the time step is 0.0089 μs (CFL = 0.0938). The absorbing layer in MDM/MMDM is enabled where $\gamma_{max}$ is 0.5 and $\alpha$ is 0.05. Pressure fields on three orthogonal planes are shown in Fig. 6. The three planes intersect at the natural focal point, i.e., the center point of the near-field distance plane (z=a$^2$/λ=34.3 mm, where *a* is the transducer radius). Figs. 6(a)-(c) show results from k-Wave; Figs. 6(d)-(f) show results from MDM; Figs. 6(g)-(i) are generated by MMDM; Figs. 6(j)-(l) are generated by MMDM4. Figs. 6(m)-(o) display the pressure differences between k-Wave and MMDM4 in all three planes. The L2-norm errors calculated within a cuboid region highlighted in Fig. 5(a) (25.5000 mm × 25.5000 mm × 42.6429 mm) are 0.3433 for the MDM, 0.3199 for the MMDM, and 0.1773 for the MMDM4, respectively.

**IV. Discussion**

We have investigated the accuracy of the MMDM for modeling linear/nonlinear wave propagation in strongly heterogeneous media. It is found that with phase correction, amplitude compensation, as well as the addition of reflections, the MMDM is significantly more accurate than the original MDM for the cases tested in this study. The MMDM, in a way, is similar to the INCS method[7, 8] in that both methods aim to solve an integral equation and treat the perturbation to the incident field as a contrast source. These two methods, however, also differ in many important aspects. For instance, the MMDM solves the integral equation using a marching scheme whereas the INCS method uses iterative approaches (e.g., the Neumann scheme). For 3D (x,y,z) problems, the INCS method operates on 4D (x,y,z,t) matrices while the MMDM involves 3D (x,y,t) matrices for transient problems or 2D (x,y) matrices for steady-state problems. The INCS method is more accurate in computing the multiple scattering provided that the iterative scheme is robust, whereas for the MMDM, the integral equation does not account for the multiple scattering and it has to be calculated in an *ad hoc* manner, whose accuracy hinges on 1D approximation which we will discuss at length below.

Some deviations in terms of the pressure amplitude between the MMDM and k-Wave results can be observed in the skull case (e.g., Fig. 3). This is likely due to the fact that the amplitude compensation introduced in the MMDM is based on the 1D assumption, though there is also the possibility that k-Wave results are less accurate for a complicated structure like the skull[29], given the relatively low spatial resolution used in this case (limited by CT scan). This could also explain why the amplitude deviation is less visible in the layered medium case, as the layer has a more regular shape and therefore the 1D assumption is more applicable and k-Wave results are also potentially more accurate in this case. To confirm this, we investigate a case where the layer is tilted at an angle of 11° instead of being normal to the beam direction (Fig. 7(a)). The density and speed of sound contrast are kept at 2.0. The time domain waveforms recorded at the focal point of the transducer are compared in Fig. 7(b). The L2-norm error is 1.6360 for MDM and is 0.4435 for MMDM4. In this case, larger amplitude differences are observed, while the phase correction is still robust although the 1D assumption breaks down in this case. We have also studied the L2 norm errors with different tilt angles. As shown in Fig. 7(c), the error expectedly grows as the title angle increases. While it is not clear how the values of these errors can be generalized to other cases with different ultrasound beams, layer thicknesses and source-layer distances, the error does become rather significant (above 0.7) in this specific case, when the tilt angle exceeds 20°. It should also be pointed out that, at large angles, acoustic wave models become intrinsically unsuitable for modeling transcranial ultrasound due to its inability to consider mode conversion. Another scenario where the 1D assumption could break down is when the wave field is strongly diverging (e.g., a spherical or cylindrical wave). This, however, is less relevant to therapeutic ultrasound and therefore is not discussed here.

Multiple reflections have been studied as a means to improve the model. Figure 2(b), for example, suggests that the second-order and fourth-order reflections can be accurately modeled. Figures 4 and 6 compare the spatial pressure distribution for the skull and skull-mimicking cases, simulated with k-Wave, MDM, MMDM and MMDM4. When reflections are included in the simulation, the accuracy is both visibly and quantitatively improved. For example, Fig. 4(d) indicates that the focal size is more precisely predicted and the interference pattern due to waves traveling in opposite directions is clearly exhibited.

The numerical implementation throughout this study is based on MATLAB 2018a (The MathWorks Inc., Natick, MA) on a 64-bit operating system with a 12-core 3.00-GHz Intel Xeon (R) Gold 6136 CPU (Intel Corp., Santa Clara, CA) processor and 192 GB of RAM. To simulate transient wave propagation using the 2D skull, the MMDM takes about 85 seconds, MMDM4 takes about 345 seconds and k-Wave takes about 650 seconds, given a computational domain that is about 120 mm by 60 mm and with the same spatial resolution for all three methods. For generating the results shown in Fig. (4), however, the MMDM only takes about 1.80 seconds, the MMDM4 takes about 5.7 seconds and k-Wave takes around 1200 seconds. Even if the CFL number is increased to 0.3 (the default number in k-Wave), k-Wave would still take a considerably longer time to produce the steady-state result. The fast speed of the MMDM/MMDM4 can be attributed to the fact that this method computes the acoustic field directly at the center frequency[14], therefore in this case reducing a 3D (x, y, t) problem to a 2D (x, y) problem that can be solved more rapidly with less computer memory. Though the MMDM is computationally efficient, the computation time inevitably increases when high-order reflections are considered. Thus, it is necessary to conduct a convergence study on multiple reflections: how many reflections are necessary for achieving satisfactory results? To this end, we first examine the steady state results for the 2D skull model with the MMDM. The L2-norm errors are calculated using MMDM50 result as the benchmark solution and the errors are plotted in Fig. 8(a). The result is considered converged when the L2-norm error is less than 0.02. It can be seen that the result is indeed converging, indicating a less significant role of higher-order reflections. The fourth-order reflection is required for the lossless skull simulation to attain sufficiently accurate results. For a more realistic simulation, absorption, which can be deduced by the density[30], is added to the skull. The absorption coefficient varies from $0.005 \text{ dB} \cdot \text{MHz}^{-y} \text{cm}^{-1}$ to $23.45 \text{ dB} \cdot \text{MHz}^{-y} \text{cm}^{-1}$. The power law exponent is 2.0 in this case, as there is no well-established data on the power law exponent for skulls. The result is seen to converge faster with the consideration of absorption, and in this case, the second-order reflection is sufficient. The results of k-Wave are not provided for the lossy skull case since k-Wave is less accurate when large absorption values are considered[31].

Two additional questions naturally arise: is it necessary to apply the corrections for soft tissue where the heterogeneities are relatively weak? Is it necessary to consider multiple reflections in soft tissue? To answer this question, a tissue map is considered as illustrated in Fig. 9(a). The acoustical properties for different tissue parts are listed in Table I. The excitation pressure magnitude is 1 Pa and the center frequency is 0.7 MHz. The transducer focal length is 49.5 mm and its diameter is 33.2 mm, corresponding to an *F* number of 1.5. The lossless case is first considered. The 2D pressure distributions obtained by k-Wave, MDM, and MDM2 are shown in Figs. 9(b)-(d). The axial pressure distributions along x=0 are shown in Fig. 9(e) for k-Wave, MDM, MMDM, and MDM2, respectively. The L2-norm error calculated using the whole domain is 0.1432 for the MDM and is 0.1403 for the MMDM (the 2D pressure distribution calculated by the MMDM is not shown in this paper). Thus, the corrections in phase and amplitude do not significantly improve the MDM in this case. The L2-norm error, on the other hand, is 0.1181 for the MDM2, indicating that in this case the inclusion of reflections is in fact more important than the correction. While these results suggest that it may not be necessary to apply corrections or even reflections to the MDM for soft tissue, this conclusion should be scrutinized for problems involving considerably larger computational domains, since the phase and amplitude errors grow along the wave propagation direction in the MDM. The convergence study for the lossless soft tissue and lossy soft tissue are also carried out and the results are shown in Fig. 8 with the MMDM50 results as the benchmark. It can be concluded that in this soft tissue case, up to the second-order reflection could be sufficient to obtain converged results (L2-norm errors being smaller than 0.02). Please note that the error 0.02 was chosen empirically as the criterion. If a different error was chosen, the conclusion regarding the maximum order of reflection could change. For example, if 0.01 were to be used, then convergence will require up to the $3^{rd}$ order reflection. Therefore, ultimately the reflection should be modeled based on the specific need of the user.

We have also studied the order of convergence *p* by using the following equation

$$error = C\frac{1}{n^q}, \qquad (19)$$

where $q$ is the order of convergence, $C$ is the asymptotic constant, and $n$ is the order of the reflection. Through fitting the curves, the convergence order for the skull without attenuation is found to be 2.01 and is 3.5 for the case with attenuation. The convergence order for the soft tissue without attenuation is 4.639 and is 4.71 for the case with attenuation. Additionally, the computation time is a linear function of the reflection order $n$. For example, the computation times for the 2D skull with different $n$ are shown in Fig. 8(b).

Finally, while the previous L2-norm errors are computed using the total acoustic field, we have also calculated the errors using only the scattering field. The scattering is computed by first conducting the simulation using the homogeneous medium (incident field) and then using the heterogeneous medium (total field). Subsequently, we calculate the difference between the two results to extract the scattering field. These errors, along with previous ones, are summarized in Tables II and III. One thing to note is that, the errors calculated using only the scattering field for the soft tissue is rather large, even with multiple scattering taken into consideration, e.g., 0.6410 for MDM2. This is because the scattering field in this case is rather weak and therefore is more susceptible to errors.

## V. Conclusion

In this paper, phase correction and amplitude compensation are proposed and implemented in the MDM so that the algorithm is more suited to modeling wave propagation in strongly heterogeneous media. The resulting model, i.e., the MMDM, is evaluated by studying three cases with strong speed of sound and density contrasts. Simulation results show that the MMDM is markedly more accurate in terms of predicting the phase and amplitude of the waveform, provided that the ultrasound beam is approximately perpendicular to the phase aberrating layer as demonstrated by Fig. 7(c). It is also shown that reflections can be added to the MMDM to further improve the accuracy of the model. Convergence studies show that the second-order reflection is sufficient for soft tissue and lossy skull simulations. While the computation time increases with the addition of reflections, the MMDM is still computationally efficient when used to predict the steady-state wave field at specific frequencies of interest. In the future, the MMDM can be coupled with the bioheat transfer equation to estimate temperature elevation in tissue. Backward propagation can also be investigated for applications such as phase correction and photoacoustic tomography in heterogeneous media.

**Acknowledgement**

This work was supported by the U.S. National Institutes of Health (Grant No. R01EB025205).

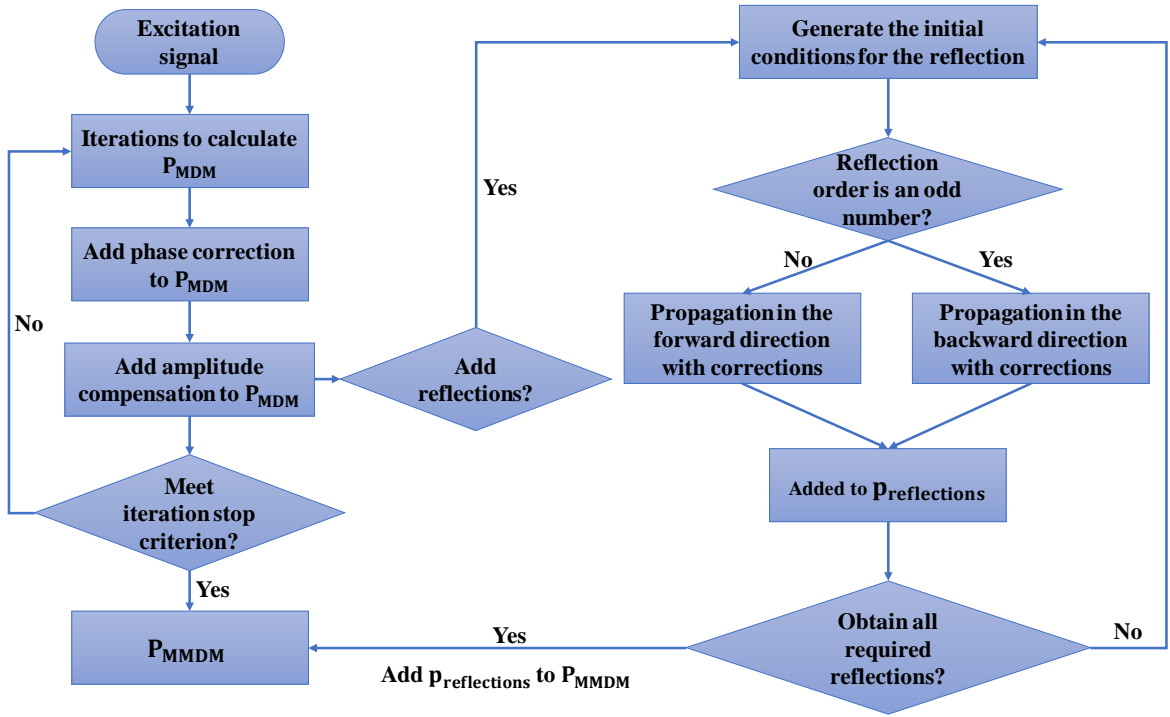

Figure 1. A flowchart illustrates the scheme to add corrections and reflections to the MDM. At each iteration step, both the phase correction and amplitude compensation are added to the MDM result. For each propagation, the reflected wave field on each plane is calculated and stored. These wave fields then propagate with both phase and amplitude corrections in the forward/backward direction. The total pressure field is finally obtained by superposing the transmission and reflection wave fields.

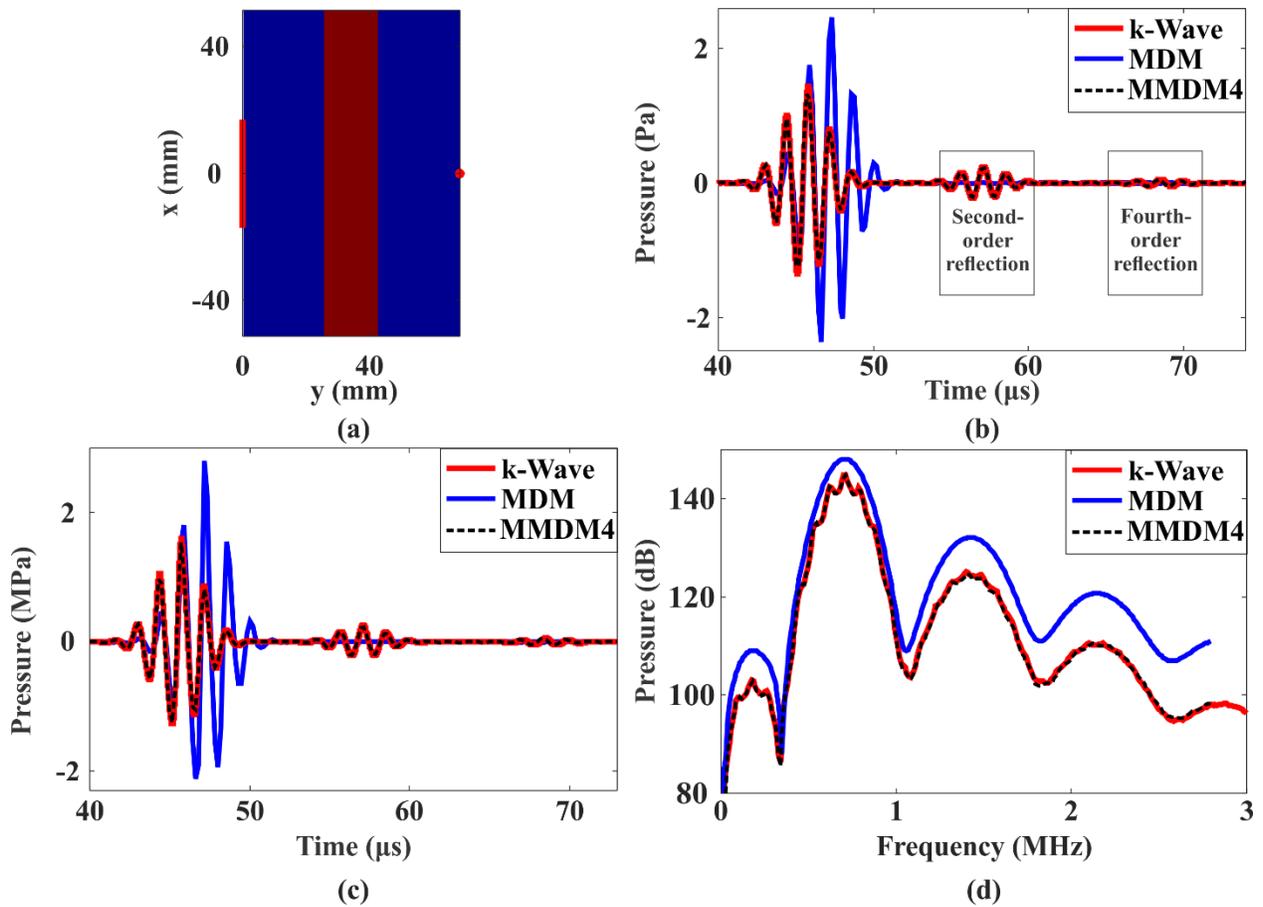

Figure 2. (a) A 2D layered medium. The red line indicates the position of the phased array transducer and the red dot indicates the position of the transducer focus. (b) Waveforms recorded at the geometrical focus of the transducer. The results simulated by k-Wave, MDM and MMDM4 are compared when the medium is linear. (c) Time-domain and (d) frequency domain results at the geometrical focus of the transducer. The results simulated by k-Wave, MDM and MMDM4 are compared when the nonlinear effect is considered.

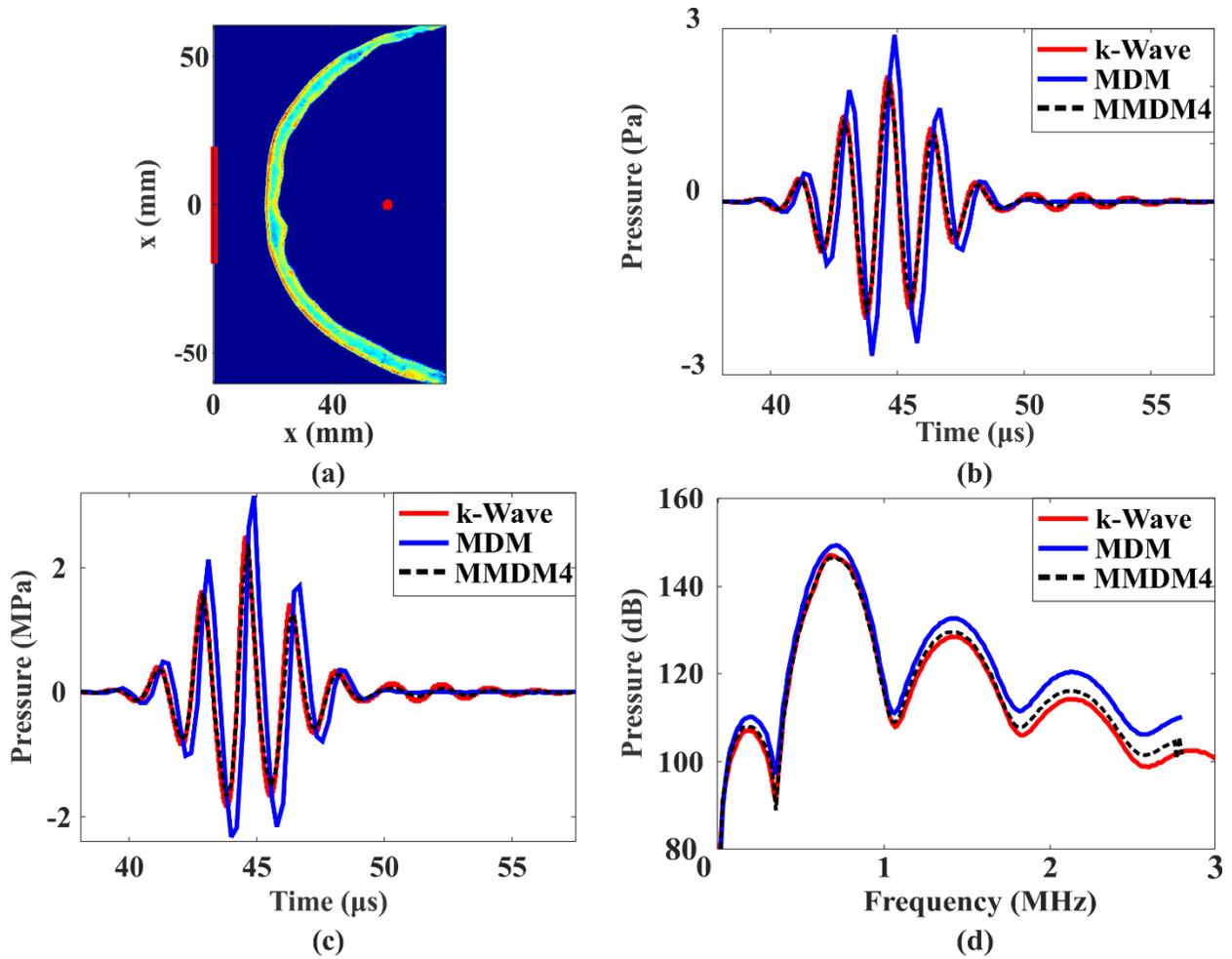

Figure 3. (a) A 2D skull model. The red line on the left indicates the array position and the red dot is the geometrical focus. (b) Waveforms recorded at the geometrical focus of the transducer. The results simulated by k-Wave, MDM and MMDM4 are compared when the medium is linear. (c) Time-domain and (d) frequency domain results at the geometrical focus of the transducer. The results simulated by k-Wave, MDM and MMDM4 are compared when the nonlinear effect is considered.

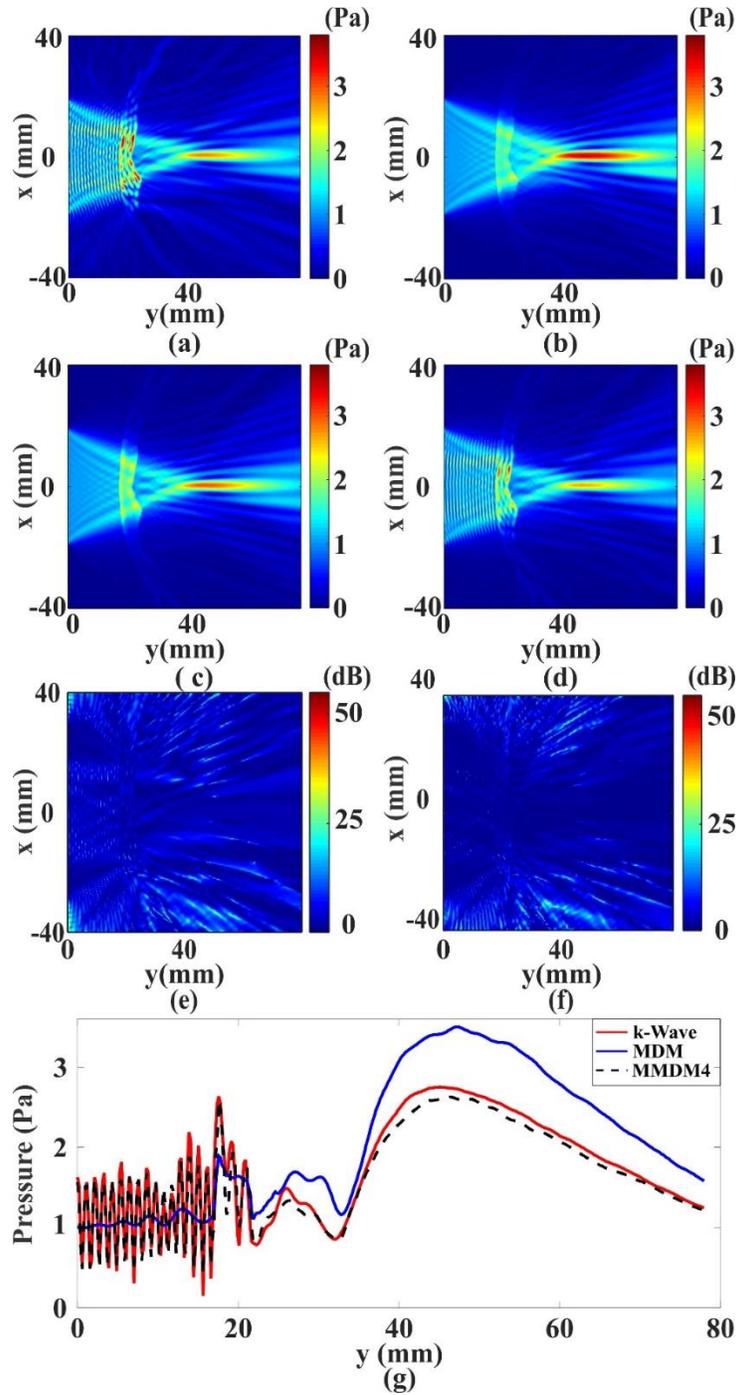

Figure 4. Spatial pressure distributions for the 2D skull case simulated with (a) k-Wave, (b) MDM, (c) MMDM and (d) MMDM4. (e) The difference in dB between MDM and k-Wave. (f) The difference in dB between MMDM4 and k-Wave. (g) Comparison for the axial pressure distribution along x=0.

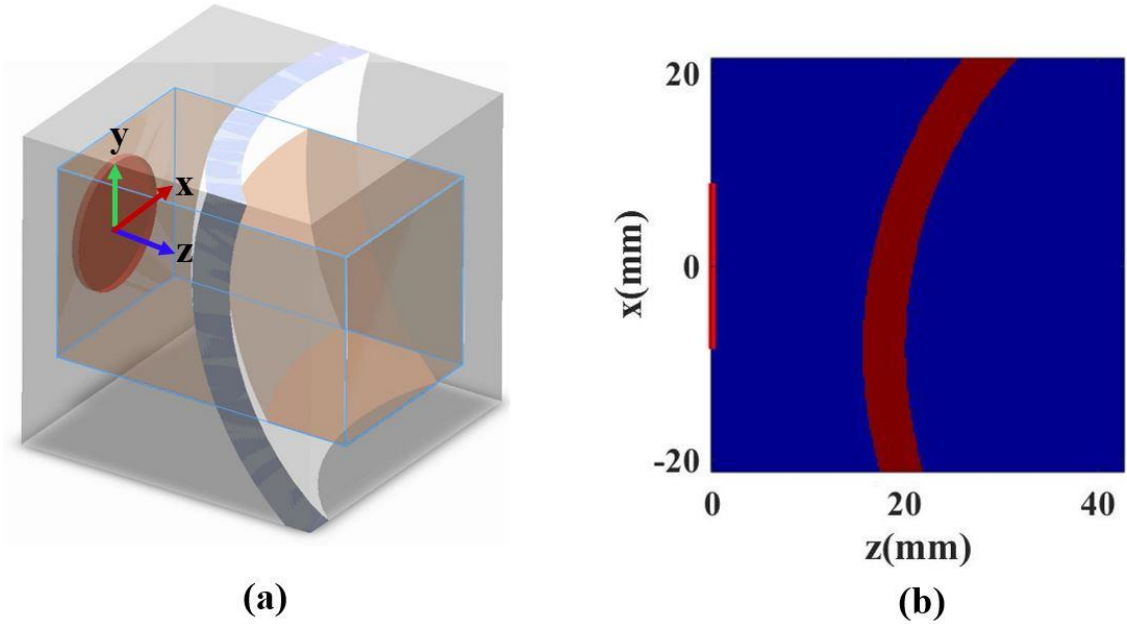

Figure 5. (a) Illustration of the 3D skull-mimicking model. The red disc indicates the size and position of the unfocused transducer. The inner cuboid is the domain where errors are estimated. (b) A cross-sectional view of the 3D heterogeneous medium on the x-z plane (y=0). The red line shows the position of the unfocused transducer.

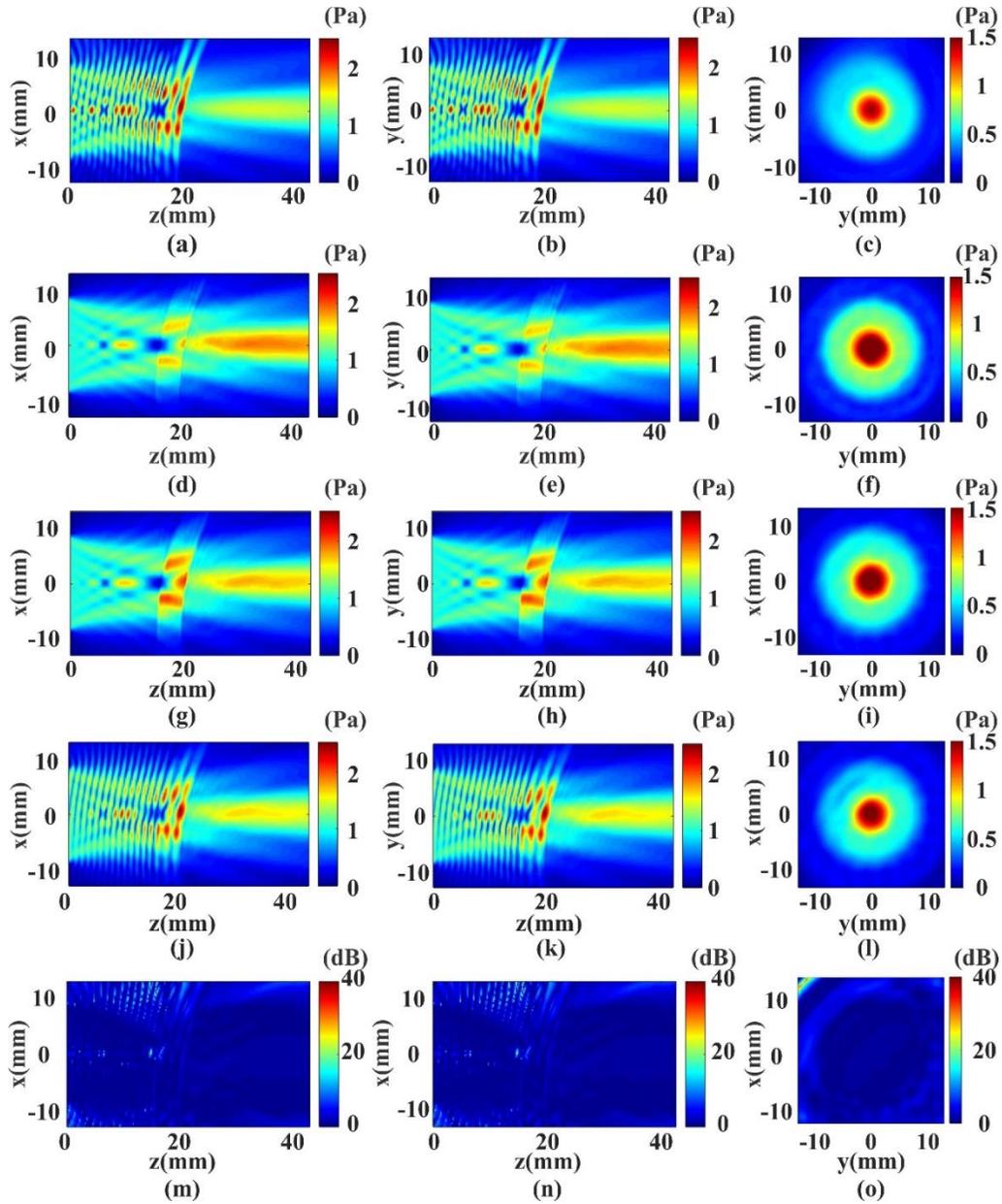

Figure 6. Pressure fields for the 3D model. (a)-(c) are generated by k-Wave; (d)-(f) are generated by MDM; (g)-(i) are generated by MMDM; (j)-(l) are generated by MMDM4. The first column ((a), (d), (g), (j)) displays the pressure field on the x-z plane at y=0. The second column ((b), (e), (h), (k)) displays the pressure field on the y-z plane at x=0. The third column displays the pressure field on the x-y plane at the near field distance. (m)-(o) displays the pressure difference between k-Wave and MMDM4.

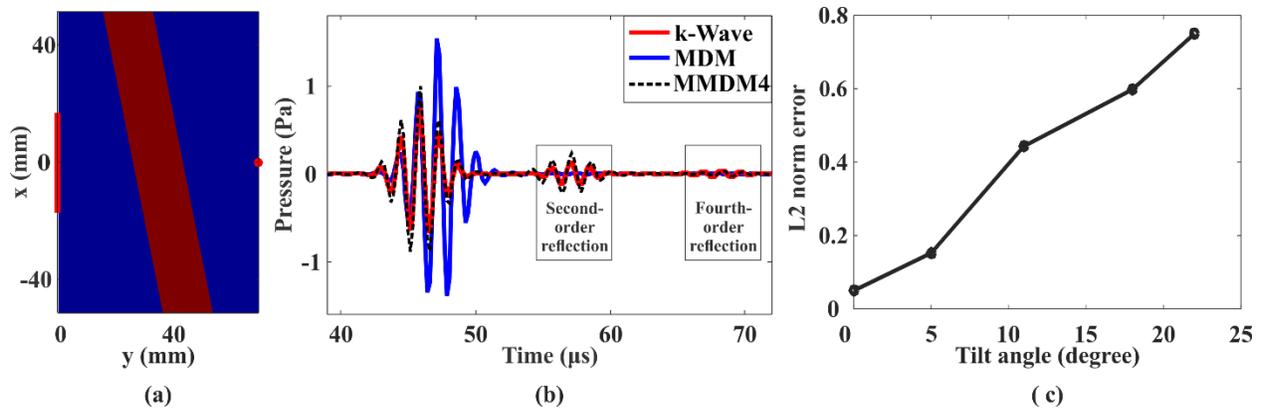

Figure 7. (a) A 2D oblique-layered media. The red line indicates the position of the phased array transducer and the red dot indicates the position of the transducer focus. (b) Comparison of the waveforms at the geometrical focus of the transducer simulated with k-Wave, MDM and MMDM4. (c) L2-norm error as a function of the tilt angle.

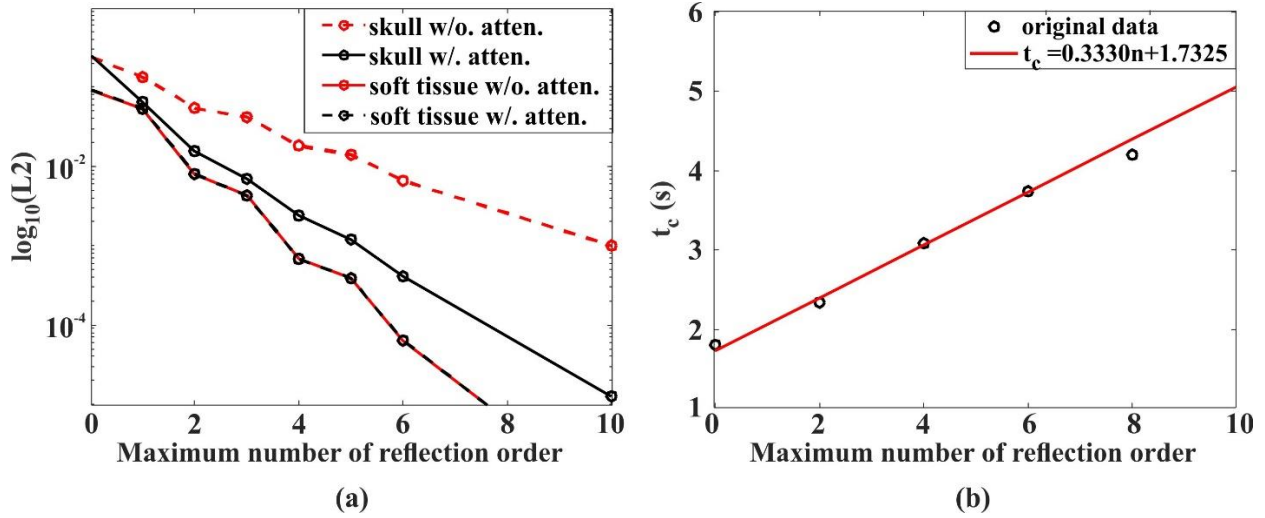

Figure 8. (a) Reflection convergence study of the MMDM for the 2D skull and soft tissue cases with and without the attenuation: L2-norm error vs. the order of reflection. (b) Computational time vs. the order of reflection for the 2D skull simulation.

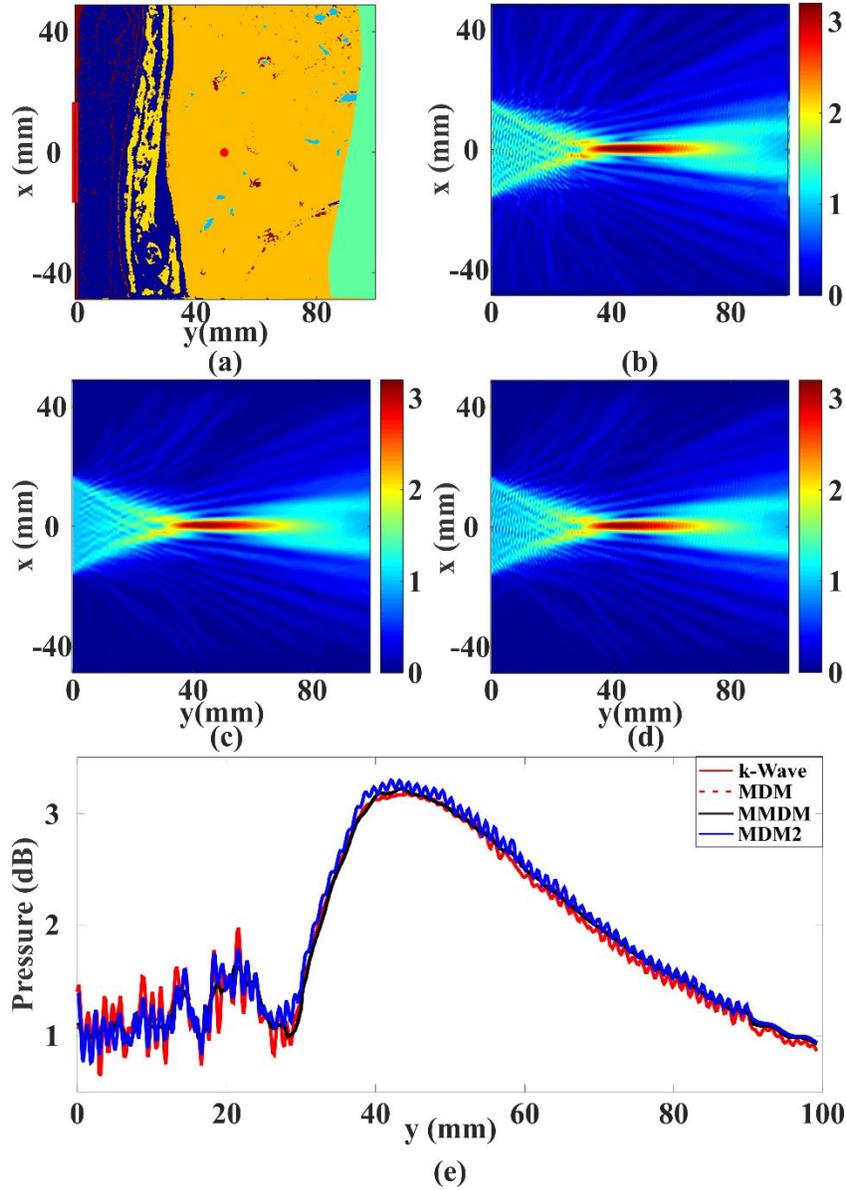

Figure 9. (a) A 2D human tissue map. The superficial layers from the left to the right denote connective tissue (red), fat (dark blue) with embedded connective tissue (red), muscle (yellow), liver (orange) and tissue (green). Blood (light blue) is inside the liver. The red line on the left boundary indicates the array position. The red dot is the geometrical focus. Spatial pressure distributions simulated with (b) k-Wave, (c) MDM, and (d) MDM2 are shown. (e) Comparison of the axial pressure distributions along the beam axis.

TABLE I

TISSUE ACOUSTICAL PROPERTIES

| | Nonlin. Coef. | Speed of sound m/s | Density kg/m$^3$ | Atten. Coef. @ 1 MHz (dB/cm) | Power law exponent y |
|---|---|---|---|---|---|
| connective | 5.0 | 1613 | 1120 | 1.57 | 1.1 |
| fat | 5.8 | 1478 | 950 | 0.48 | 1.1 |
| muscle | 5.5 | 1547 | 1050 | 1.09 | 1.1 |
| liver | 4.3 | 1595 | 1060 | 0.5 | 1.2 |
| blood | 4.05 | 1584 | 1060 | 0.2 | 2.0 |
| tissue | 5.5 | 1540 | 1000 | 0.5 | 1.1 |

TABLE II

L2 NORM ERRORS FOR THE TIME-DOMAIN RESULTS

|  | MDM | MMDM4 | MDM* | MMDM4* |
|---|---|---|---|---|
| Linear parallel layer | 1.2137 | 0.0509 | 0.5743 | 0.0412 |
| Nonlinear parallel layer | 1.2234 | 0.0579 | 0.5715 | 0.0695 |
| Linear skull | 1.0716 | 0.1120 | 0.8201 | 0.0671 |
| Nonlinear skull | 1.0967 | 0.1261 | 0.8105 | 0.0842 |
| Oblique layer (11°) | 1.6360 | 0.4435 | 0.4208 | 0.1227 |

* The error is calculated by using only the scattering field

TABLE III

L2 NORM ERRORS FOR THE FREQUENCY-DOMAIN RESULTS

|  | MDM | MMDM | MMDM4 for skull and MDM2 for soft tissue |
| --- | --- | --- | --- |
| 2D skull | 0.3374 | 0.2839 | 0.1717 |
| Soft tissue | 0.1432 | 0.1403 | 0.1181 |
| 3D skull | 0.3433 | 0.3199 | 0.1773 |
| 2D skull* | 0.7816 | 0.6421 | 0.3622 |
| Soft tissue* | 0.7568 | 0.7412 | 0.6410 |
| 3D skull* | 0.9333 | 0.8539 | 0.4525 |

* The error is calculated by using only the scattering field